\documentclass[11pt,final,letterpaper,onecolumn,compsoc]{IEEETran}
\usepackage{graphicx}
\usepackage{amsmath,alltt,color,url}
\usepackage{mathtools,stmaryrd}
\usepackage{fancyvrb,relsize}
\begin{document}
\title{Industrial Strength Formal Using Abstractions}
\author{Ashish Darbari and Iain Singleton\\Imagination Technologies\\Homepark Industrial Estate\\Kings Langley, WD4 8LZ, UK}

\maketitle 

\newcommand{\hh}[1]{\#\#{#1}}
\renewcommand{\mapsto}{\texttt{|->}}
\renewcommand{\Mapsto}{\texttt{|=>}}

\begin{abstract}
Verification of concurrent systems with thousands of multiple threads and transactions is a challenging problem not just for simulation or emulation but also for formal. To get designs to work correctly and provide optimal PPA the designers often use complex optimizations requiring sharing of multiple resources amongst active threads and transactions using FIFOs, stallers, pipelining, out-of-order scheduling, and complex layered arbitration. This is true of most non-trivial designs irrespective of a specific application domain such as CPU, GPU or communication. At the outset a lot of these application domains look diverse and complex; however at some level of detail all of these employ common design principles of sequencing, load balancing, arbitration and hazard prevention. We present in this paper a key abstraction based methodology for verifying ordering correctness and arbitration across a range of designs which are derived from different application domains. We show how by using transaction tracking abstractions and supporting them with invariants one can not only find deep corner case bugs in sequentially very deep designs; we can also build exhaustive proofs to prove the absence of critical bugs such as deadlock, starvation, and loss of data integrity. We present experimental results on a range of different designs and show that on some of the designs such as FIFOs we can verify over 100 different types of FIFOs for ordering correctness using a single assertion in a single testbench.  We also show how the methodology of FIFO verification can be adapted to verify over half-a-dozen different types of arbiters including a very complex memory subsystem arbiter. We then verify a multi-clocked synchronizer and a packet based design from a networking domain using the same abstraction as used in FIFO verification. The results demonstrate that by using our methodology one can build scalable formal verification testbenches which can be both reusable and compact, and can be used to find deep corner case bugs as well as establish exhaustively the absence of bugs through proofs.
\end{abstract}
\section{Introduction}
Several industry estimates point to an increase in costs of hardware verification. A notable study done recently~\cite{Foster2014} uncovers that ``the average total project time spent in verification in 2014 was 57\%", with 20-25\% projects reported to have consumed as much as 60-70\% time in verification and about 10\% of the projects in excess of 80\%. Hardware designs are shrinking in die size and are getting more energy efficient, however the feature set of requirements these designs need to have is increasing --- specially as the use models are getting more diverse and pervasive from internet-of-things (IoT) to automotive, avionics and medical electronics. Meeting schedules on time and providing rigorously verified systems is becoming a daunting challenge which can only be addressed through rigorous verification techniques grounded in formal methods. Even though test based techniques based in simulation and emulation are an integral part of any verification plan, they often are not the best choice as it takes longer to implement these and also longer to run. It is not feasible to continue to extend the time taken for regression testing, even with the use of emulation and large server farms. Also, neither simulation nor emulation can ever provide the much needed rigor and often coverage is the only way to sign-off --- which for domains such as automotive may not be necessarily adequate.

Verification of concurrent systems with thousands of multiple threads and transactions is a challenging problem not just for simulation or emulation but also for formal. To get designs to work correctly and provide optimal power, performance and area (PPA) the designers often use complex optimizations requiring sharing of multiple resources amongst active threads and transactions using FIFOs, stallers, pipelining, out-of-order scheduling, and complex layered arbitration. This is true of most non-trivial designs we build irrespective of a specific application domain such as CPU, GPU or communication. At the outset a lot of these application domains look diverse and complex; however at some level of detail all of these employ common design principles of sequencing, load balancing, arbitration and hazard prevention. 

In this paper we present an abstraction based methodology for verifying designs that require ordering correctness and arbitration which are derived from different application domains. We introduce a family of abstractions built around \textit{tracking transactions} and \textit{splitting scenarios} and show how by using these one can not only find corner case bugs in sequentially deep designs; we can also build exhaustive proofs to mathematically prove the absence of critical bugs such as deadlock, starvation, and loss of data integrity. 
\subsection{Organization of the paper}
In the following section we present some context by summarizing background work in the field of problem reduction, abstraction and model checking. In Section~\ref{sva}, we present a brief introduction to SystemVerilog assertions (SVA) that we have used in our work. We then present our first case study of a FIFO in Section~\ref{fifo} and introduce our first key abstractions in that section. In Section~\ref{simplearb}, we show how the fundamental abstractions introduced for FIFO verification are used in verifying a simple arbiter. In Section~\ref{mem_arbiter} we introduce a new abstraction and show how this one together with the abstractions used for FIFO verification is used to verify a more complex memory subsystem arbiter. In Section~\ref{sync}, we show how multi-clocked synchronizers are verified for data integrity and finally our last case study of packet based design verification is discussed in Section~\ref{packet} where we show how a packet based design where packets can be dropped, shrunk or grown can be verified using the abstraction used for verifying FIFOs. We then provide a detailed comparison of our work in Section~\ref{related} with other closely related pieces of work and finally conclude in Section~\ref{concl} with our key findings and contributions.
\section{Background Work}
Abstraction~\cite{Cousot:1977:AIU:512950.512973,DBLP:journals/toplas/ClarkeGL94} has been a classic technique used to cope with verification complexity for nearly 40 years. Since the invention of model checking~\cite{Clarke:1981:DSS:648063.747438,Clarke:1986:AVF:5397.5399,Queille:1982:SVC:647325.721668} abstraction has been deployed in numerous forms in verification of hardware~\cite{DBLP:books/cu/Melham93,jones-dandt-01,mcmillan00methodology,DBLP:conf/fmcad/AggarwalCKS11} as well as software~\cite{DBLP:conf/pldi/BallMMR01}. The most challenging aspect in deploying abstraction successfully is to devise one in the first place that is relevant to the target application and to establish that the abstraction used is \textit{appropriate }and \textit{safe.} Appropriate means the abstraction should not abstract away so much of the underlying design that one starts getting spurious errors (failures that aren't reflective of real bugs). The abstraction is safe if it does not abstract away so much of the design to get proofs to exhaustively complete on an unrealistically simplified design and ends up missing bugs in the original un-abstracted design. It is well known that if one is able to build a proof on an abstracted design model that proof holds for the original design as well~\cite{DBLP:journals/toplas/ClarkeGL94}. However, one must keep the delicate nature of abstraction in mind to ensure that we do not end up computing abstractions that produce the above mentioned problems. Establishing that the abstraction is a safe one and will not prevent one from unearthing some bugs that may exist in the original unabstracted design is a delicate art. 

Often it may be possible to build a proof that these abstractions are safe by theoretically establishing a Galois connection~\cite{Cousot:1977:AIU:512950.512973,Chou:1999:MFF:647768.733788}. The absence of a Galois  connection however does not limit the practical use of abstraction~\cite{mcmillan00methodology,DBLP:conf/fmcad/AggarwalCKS11}. We do however appreciate that we need to check that abstractions we use are indeed safe. In our case we make use of mutation based testing to establish that abstractions are safe and we have not missed bugs in the original design through our abstractions. This is our preferred choice of method currently in establishing safety of our abstractions.

The use of symmetry based reduction has been significant in the verification of cache-coherency protocols~\cite{Ip:1996:BVT:235947.235950,Clarke1996,mcmillan00methodology}, in the verification of hardware designs~\cite{DBLP:conf/fmcad/Darbari06} and in verification of telecommunication protocols~\cite{DBLP:journals/csur/MillerDC06}. McMillan~\cite{mcmillan00methodology} notably used assume guarantee reasoning to verify hardware designs with loops. The idea here was to break the circularity to assume one half of the design works to prove the other and then assume the half just proven to prove the other half of the loop.  
Melham et al~\cite{DBLP:conf/fmcad/MelhamJ02} showed how symbolic indexing can be used to efficiently encode numerous cases required to verify some hardware designs. 

Data independence~\cite{Wolper:1986:EIP:512644.512661,DBLP:journals/entcs/BenalycherifM09,DBLP:conf/fmcad/AggarwalCKS11} is a very powerful technique in verifying systems where the data simply moves within the design or a program without changing its value. Its correspondence with symmetry~\cite{Lazić2000} has been shown to be a very effective way of applying data-independence techniques on systems with symmetry. 
\section{A Brief Introduction to SystemVerilog Assertions}\label{sva}
We assume familiarity with Verilog programming language. Though most of our methodology is grounded in SystemVerilog Assertions~\cite{cohen2005systemverilog}, it does not assume familiarity with the full set of SystemVerilog (SV). On the contrary, we only ever use a tiny subset of SV and employ mostly Verilog for modelling glue logic and wrap that with SV properties. Properties in SV form the core of what is known as SystemVerilog Assertions (SVA). SVA and and Property Specification Language (PSL)~\cite{Geist2003} are both languages used for writing assertions and are mostly used with model checkers although in some cases they can also be employed to capture checkers for sequential equivalence checking as well as dynamic simulation based checking.

Both these assertion languages are used in industry and are based on Linear Temporal Logic~\cite{Pnueli:1977:TLP:1382431.1382534}. SVA however has been adopted a lot more than PSL and is certainly the standard assertion language used in Imagination Technologies. These languages come with a rich repertoire for modelling Boolean expressions and also complex temporal expressions called sequences built by applying temporal operators such as {\it single cycle implication}, {\it next cycle implication}, and {\it repetition operators} to model consecutive or non-consecutive repetition as well as {\it liveness properties}. As much as these assertion languages are well endowed with features modeling time-dependent behaviour it is non-trivial to model complex time-dependent behaviour for end-to-end checks in these languages. It is also not clear how easy it is to reuse these properties for other similar designs just by using SVA without any supporting modeling. 

The usage of the word assertions in SVA is somewhat legacy as clearly assertions only make a part of the SVA and in fact it is the properties that are the building blocks of the formal layer of SVA. One can express combinational or sequential behaviour by defining properties in SVA using a very rich set of syntactic sugar that SVA provides and then one can assert that a property should always hold of the outputs or (in some cases of internal state) of RTL design at all times using the keyword $\mathtt{assert\;property}$. To model realistic behaviour we often need to capture real world constraints that will restrict the behaviour of the inputs of the design to only a subset of interesting attributes. This is accomplished by using the keyword  $\mathtt{assume\;property}$ in SVA which states that at all times, the behaviour on an input of the design is always true. This forces an automated formal tool such as a model checker or an equivalence checker to use this to force some of the inputs of the design to always have the behaviour captured by an assume property in order to prove that the required behaviour at all times expressed by assert property is valid. 

In many situations it is desirable that some behaviours are not always true but may be true some time. For example, one would not want a FIFO to be always empty, or always full; but may want to check that a FIFO does become empty or full sometime. This is captured by using a construct in SVA called a $\mathtt{cover\; property}$.

We do not have space to provide a detailed introduction to SVA here but we would like to introduce just those preliminaries of SVA here that would be needed to appreciate our modelling and case studies in the later section. We show concisely in a table the essential SVA we need to use in our paper.
\begin{table}[!h]
{\begin{tabular}{|l|l|l|l|}
\hline
\scriptsize{\texttt{expr1 \&\& expr2}}  & \scriptsize{{Logical `and' of \texttt{expr1} and \texttt{expr2}}}  
 & \scriptsize{\texttt{A $\mapsto$ B}}  & \scriptsize{{When \texttt{A} holds \texttt{B} must hold in same cycle}}\\
\hline
\scriptsize{\texttt{expr1 || expr2}}    & \scriptsize{{Logical `or'  of \texttt{expr1} and \texttt{expr2}}}
& \scriptsize{\texttt{A $\Mapsto$ B}}   & \scriptsize{{When \texttt{A} holds \texttt{B} must hold in next cycle}}
\\ \hline
\scriptsize{\texttt{expr1 \& expr2}}    & \scriptsize{{Bitwise `and' of \texttt{expr1} and \texttt{expr2}}} 
& \scriptsize{\texttt{\hh N\, expr}}    & \scriptsize{{\texttt{expr} holds after \texttt{N} cycles}}
\\ \hline
\scriptsize{\texttt{expr1 | expr2}}     & \scriptsize{{Bitwise `or'  of \texttt{expr1} and \texttt{expr2}}} 
& \scriptsize{\texttt{\hh[0:\$] expr}}  & \scriptsize{{The \texttt{expr} holds eventually}}
\\ \hline
\scriptsize{\texttt{!expr}}             & \scriptsize{{Logical negation of \texttt{expr}}} 
& \scriptsize{\texttt{\$stable(expr)}}  & \scriptsize{{The \texttt{expr} value doesn't change with time}}
\\ \hline
\end{tabular}}\\
\label{sva_tbl}
\caption{\scriptsize{A quick primer on SVA syntax showing some of the commonly used operators in SVA. Combinational fragment of SVA is shown in the left column and temporal fragment in the right column.}}
\end{table}
\normalsize
One important aspect to remember when using properties in SVA is that they are mostly clocked when they are meant to capture time based behaviour. This is so that the model checkers do not have to check or assume the behaviour when the clock is not toggling and this makes the model checking very efficient. In fact it is not a requirement that one has to use a clock, any event can be used as long as it is has an edge like clocks have positive ($\mathtt{posedge}$) and negative ($\mathtt{negedge}$). 

The template for modelling properties then becomes:\\
\begin{Verbatim}[fontsize=\relsize{-2}, frame=single]
[assert|assume|cover] property (@(posedge clock) ...);
\end{Verbatim}

where one can use one of the keywords between $\mathtt{assert}$, $\mathtt{assume}$ and $\mathtt{cover}$ and the three dots denote a combinational expression or a temporal expression such as the ones shown in Table~\ref{sva_tbl}. 

If we do not want to explicitly use the edge qualifier $\mathtt{@(posedge\; clock)}$ one can define default clocking blocks and then we no longer have to qualify explicitly the clock edge in the property. However, we must be careful when verifying designs with multiple clocks as is the case in Section~\ref{sync} where we do have to explicitly qualify multi-clocks and edges. However, when these are not specified explicitly in our presentation we can safely assume that we have used default clocking blocks. The default clocking in SVA for the clock $\mathtt{clk}$ in our design is specified as:
\begin{Verbatim}[fontsize=\relsize{-2}, frame=single]
default clocking cb @(posedge clk);
endclocking
\end{Verbatim}
{\textbf{A note on notations and results:} Our models used in formal are designed using a synthesizable subset of Verilog using the standard registers and wires. When we show our modelling code we will not show the declarations of wires and registers as this may become obvious from the way these signals are assigned values in Verilog. Only when it is essential to show key aspects of data structures we will explicitly show these declarations. Also, one needs a $\mathtt{begin}$ and $\mathtt{end}$ with $\mathtt{for}$ loops in Verilog but we will omit this in the presentation to save space and enhance readability. Due to the model checker not able to report CPU times when using certain solvers based, we are only able to report real time or wall-clock time for some of our case studies which is of course more pessimistic than CPU times.}
\section{The Case of a Humble FIFO}\label{fifo}
A FIFO is one of the most used components in a SoC design. Several implementations are possible but some are better than others for power and area. Though FIFOs are often classed as somewhat trivial components we have seen that in practice it is quite easy to have a bug in one of these components specially when power based optimizations are implemented. Moreover, formal verification in the presence of FIFOs is often considered difficult; in some cases infeasible to an extent that it is suggested that black-boxing is necessary. In practice, we cannot black-box them all the time as they are part of essential control flow. Moreover not verifying them exhaustively in the presence of power based optimizations is not an ideal choice. Imagination Technologies has a library of design components including FIFOs and when we investigated building a formal test bench to verify one of these we soon realized having an end-to-end check for establishing the key correctness properties was essential so we can reuse these checks regardless of the specific micro-architectural implementation of a given FIFO. Most FIFOs have to satisfy the following key properties:
\begin{enumerate}
\item [R1] No data loss
\item [R2] No data duplication
\item [R3] No reordering allowed
\item [R4] All data words accepted at the input are eventually received at the output
\end{enumerate}
\normalsize
Most practitioners will agree that establishing exhaustively that checks (R1-R4) even for moderate sized FIFOs (greater than 32 deep) is challenging for formal model checking. The reason for this is because the state of each stored element in the FIFO depends on whatever else has been stored ahead of it; enforcing sequential dependency (required for FIFO) but also making it harder for the model checkers to scale with increasing depth as this dependency causes an exponential blow up in the size of BDDs or SAT based model checking. This happens if one uses another golden FIFO design to verify a FIFO RTL using cycle-accurate sequential equivalence checking. For an N deep FIFO with W wide data words; it will have $2^{(N\,\times\,W)}$ states reachable from reset; and if one introduces another FIFO with these many states $2^{(N\,\times\,W)}$ then the model checker has to cope with a cross product of $2^{(N\,\times\,W)}\,\times\, 2^{(N\,\times\,W)}$ states which makes the problem significantly harder specially in those conditions where the depth N is a big number or when FIFOs are used as design components in a bigger design.
\subsection{Two Transaction Abstraction}
\begin{figure}
\begin{center}
\includegraphics[scale=0.38]{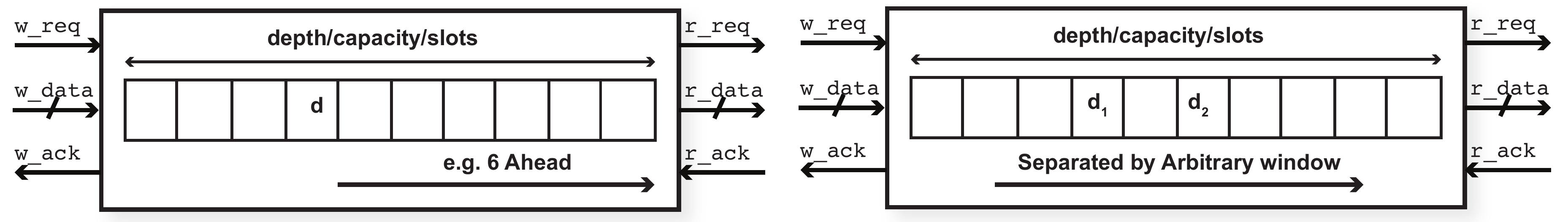}
\caption{\scriptsize{The Two Abstractions. On the left hand side is the smart tracker abstraction designed by tracking single symbolic arbitrary watched data word while on the right hand side is the two transaction abstraction showing tracking of two symbolic arbitrary watched data words.}}
\label{two_abs_fig}
\end{center}
\end{figure}

Our first abstraction~\cite{DarbariCDNLive2014} shown on the right hand side in Figure~\ref{two_abs_fig}, is known as {\it the two transaction abstraction}. If we were using theorem provers to model such checks one can write the ordering property as follows where $\mathtt{seen\_in\_before}$ and $\mathtt{seen\_out\_before}$ would be predicates. The predicate $\mathtt{seen\_in\_before}$ would be valid if $\mathtt{d1}$ arrived before $\mathtt{d2}$ on the input interface and the predicate $\mathtt{seen\_out\_before}$ would be valid if $\mathtt{d1}$ appeared on the output interface before $\mathtt{d2}$. The predicates themselves would be higher-order functions that would take a time based function $\sigma$ and the two data words $\mathtt{d1}$ and $\mathtt{d2}$ as arguments and we would then exploit induction based techniques to prove the ordering property. The induction would be carried out to establish that the property holds for all $\mathtt{d1}$ and $\mathtt{d2}$ and for all insertions of these in the design at all time points at all possible locations inside the design. \\

\;\;$\mathtt{\forall d_1 d_2.\; (d_1\, seen\_in\_before\, d_2) \Rightarrow (d_1\, seen\_out\_before\, d_2)}$ \\

In industry, the use of theorem provers is not as popular as model checkers as theorem provers cannot provide counter-examples and cannot be run automatically as model checkers do. However, theorem provers do not suffer from any scalability issues with increasing state in the design and certainly do not have any state-space explosion problems that all model checkers have. A lot of our work in this paper is motivated to offset any state-space explosion issues we have during model checking big designs.

We want our formal testbench to be simple, reusable and scalable and we decided that the best way would be to build {\it scoreboards} similar to how they are built in dynamic simulation based testbenches. However, we want to ensure that instead of tracking the state of every transaction (comprising 0s and 1s) - we only track the \textit{interesting} ones. By interesting we mean exploiting symbolic transactions i.e., the ones that are encoded using Boolean symbols and not encoded using explicit 0s and 1s. We also wanted to make use of non-determinism to choose which symbolic transactions we will track. With formal the main idea is that if one can prove a property about a non-deterministic symbolic transaction then the property is true of all transactions. A symbolic transaction is defined in terms of a non-deterministic symbolic value. First we choose a symbolic value in the design and then we define when to `start' and `stop' observing this symbolic value. Non-determinism comes by keeping this symbolic value undriven in our model and letting the model checker exercise all possible combinations of explicit 0s and 1s for this undriven symbolic value at all times. This allows us to define observation windows for this symbolic value. The non-deterministic, undriven symbolic value defined is often called as a \textit{watched data}, or a \textit{watched symbolic data}. We use these interchangeably throughout our presentation.

A state-of-the-art model checker typically combines \textit{symbolic simulation} (simulation using symbolic values), value propagation and checking using BDD~\cite{Bryant:1986:GAB:6432.6433} and SAT~\cite{Biere:1999:SMC:309847.309942} based algorithms. The model checker itself will ensure that it instantiates the watched data with all possible data values (built over 0s and 1s) and this is how exhaustive results are achieved by it. Before the model checking is started the assertion in question is logically negated before it is combined with the RTL, constraints and the model in the testbench to result in a Boolean formula. As these Boolean formulas are unfolded for every time point during model checking they are sent to the SAT solvers to compute if there is a satisfying assignment of Boolean values to variables (signals in the RTL, testbench and constraints) in this formula and if there is one then at that time point a bug is found. This bug could be in the RTL, or in the constraints or in the model of the testbench. A debug process can then identify the cause and fix the bug. 
\subsection{Building observation windows using symbolic variables}
Our abstraction method relies on defining a `start' and a `stop' event for every watched data where a start is defined as the point at which a given symbolic watched data is accepted at the input interface via some handshake signals. The stop event is defined as the point of time where the symbolic watched data leaves the design on the output interface using possibly a second set of handshake signals. 

In our method we use two symbolic watched data words to track inside the FIFO and we assert that if one came before the other then they leave in the same order in which they arrived in the design. The choice of which words would be chosen as symbolic is left entirely non-deterministic and is configured at run time by the model checking tool and it chooses to exercise all possible data values for the symbolic watched data. 
We do not control when $\mathtt{d1}$ and $\mathtt{d2}$ appear on the input interface, or how far they are separated (in time) by when they arrive on the inputs; all we do is to constrain that $\mathtt{d1}$ arrives before $\mathtt{d2}$. We then check by asserting a property that $\mathtt{d1}$ should always appear before $\mathtt{d2}$ at the output interface. The model checker tries inserting $\mathtt{d1}$ and $\mathtt{d2}$ at all possible locations within the FIFO at all possible time points and tries to prove the asserted property. If it cannot fail the property it will build an exhaustive proof. Please note that the choice of letting $\mathtt{d1}$ come before $\mathtt{d2}$ is completely arbitrary and we could have chosen it otherwise. As long as the relative chronology is intact for both the symbolic values the design is keeping all the input data correctly ordered.

The key to our model is in creating two set of registers --- one that gets set when the watched data appears on the input and another one that gets set when the sampled in watched data leaves on the output port. We also factor in the abstract model an \textit{extra} element of non-determinism to allow the sampled in register to not just  get set on the very first appearance of the watched data but also on later occurrences of the watched data as well. The reason we do so is to make sure that if there was ever an instance of multiple occurrences of data words on the input where a later occurrence is bugged (let us say its value is changed in the course of a write handshake due to an RTL bug), we would be able to catch this bug. If we do not allow this non-determinism in our abstract model then we will set the sampled in register on the very first occurrence of a watched data and will not wait for a later occurrence thereby missing the RTL bug. We accomplish this added non-determinism by using another un-driven wire in our model called $\mathtt{arbit\_window}$. We never drive this wire with any specific value and instead leave it for the model checker to assign both a 0 and a 1 to capture the non-determinism at runtime. When this wire has the value 1 and the watched data (e.g., in our case shown as $\mathtt{d1}$ in the code below) appears at the input port then the sampled in register ($\mathtt{sampled\_in\_d1}$) will get set. If what is going to be chosen as the watched data value appears at the input port and at that instant of time the wire $\mathtt{arbit\_window}$ is not 1, then the sampled in register will not be set and the model will wait for another subsequent occurrence of $\mathtt{d1}$ to arrive. The model checker itself is obliged to create another instance of an incoming data value to be the watched data value $\mathtt{d1}$ when the $\mathtt{arbit\_window}$ is 1. 

In our presentation we assume default clocking and therefore do not explicitly qualify asserts and assumes with $\mathtt{@(posedge\; clk)}$. The parameter $\mathtt{WIDTH}$ is the size of the data word. A data word of width $\mathtt{WIDTH}$ will have $\mathtt{2^{WIDTH}}$ explicit values built over 0s and 1s. 
\begin{Verbatim}[fontsize=\relsize{-2}, numbers=left, frame=single]
logic [WIDTH-1:0] d1;
assume property ($stable(d1));
assign w_hsk    = w_req && w_ack;
assign r_hsk    = r_req && r_ack;
assign seen_d1  = (w_data == d1) && w_hsk && !sampled_in_d1 && arbit_window;
always @(posedge clk or negedge rst)
if (!rst)
  sampled_in_d1 <= 1'b0;
else
  sampled_in_d1 <= seen_d1 || sampled_in_d1;
\end{Verbatim}
The $\mathtt{seen\_d1}$ wire gets set when the watched data $\mathtt{d1}$ appears on the input port on a write handshake and it has not already been sampled in before and the $\mathtt{arbit\_window}$ wire is driven to 1.  When $\mathtt{seen\_d1}$ is 1 the $\mathtt{sampled\_in\_d1}$ register is set and once set remains so. Lastly, to allow the model checker to work on a fixed but arbitrary watched data $\mathtt{d1}$ we need to explicitly keep it stable (fixed). This is done by using an $\mathtt{assume\; property}$ in SVA to tell the model checking tool that the watched data is stable (we use the $\mathtt{\$stable}$ construct in SVA). Other signals controlling the read ($\mathtt{r\_hsk}$) and write ($\mathtt{w\_hsk}$) handshake are also defined below in terms of read and write request/acknowledgment pairs respectively.

We now define a register ($\mathtt{seen\_read}$) that detects a read on the output port. This is required as our implementation has a one cycle delay between the read handshake and when any data appears at the output data port.
\begin{Verbatim}[fontsize=\relsize{-2},numbers=left,frame=single]
always @(posedge clk or negedge rst)
if (!rst)
  seen_read <= 1'b0;
else (r_hsk)
  seen_read <= 1'b1;
\end{Verbatim}
We now define the event that detects that the watched data $\mathtt{d1}$ has been read out. This happens when there has been a prior sampling in of $\mathtt{d1}$ and a read request ($\mathtt{seen\_read}$) in the previous clock cycle and the value read out on the $\mathtt{r\_data}$ port is indeed $\mathtt{d1}$.
\begin{Verbatim}[fontsize=\relsize{-2}, numbers=left, frame=single]
always @(posedge clk or negedge rst)
if (!rst)
  sampled_out_d1 <= 1'b0;
else if (sampled_in_d1 && seen_read && (r_data==d1))
  sampled_out_d1 <= 1'b1;
\end{Verbatim}
For the sake of space we are not showing the modelling code for the watched data $\mathtt{d2}$ but it is symmetric to the code shown above for $\mathtt{d1}$ and can be obtained by replacing $\mathtt{d1}$ by $\mathtt{d2}$. The only difference is that we will not include the $\mathtt{arbit\_window}$ term for computing the $\mathtt{seen\_d2}$ expression. We now define two key constraints required for ensuring we do not run into spurious failures at the time of checking. These are that the two watched data words are distinct from one another and that $\mathtt{d1}$ arrives before $\mathtt{d2}$. Again, assuming default clocking in SVA these are shown below:
\begin{Verbatim}[fontsize=\relsize{-2}, numbers=left, frame=single]
assume property (d1 != d2);
assume property (!sampled_in_d1 |-> !sampled_in_d2);
\end{Verbatim}
We now define our key ordering check which can be formalized in SVA as:\\
\begin{Verbatim}[fontsize=\relsize{-2},frame=single,label=Master Assertion]
assert property (sampled_in_d1 && sampled_in_d2 && !sampled_out_d1 |-> !sampled_out_d2);									
\end{Verbatim}
The above check will exhaustively verify all combinations of specific data word at all time points and will flag an error if the ordering check is broken. We additionally verify a liveness property which establishes that whatever went into the design comes out eventually. Even though this check does not establish any ordering it does establish that no data loss happens. Liveness properties are extremely hard to prove specially on deep FIFOs, but as we have a valid proof of the ordering property, we can assume that and then the proof of liveness property is discharged within seconds of wall-clock time by the model checker. The property is shown below. Note that we prove this only for one of the watched data words not both as this is sufficient to find any data loss bugs.
\begin{Verbatim}[fontsize=\relsize{-2},frame=single,label=Liveness]
assert property (sampled_in_d1 |-> ##[0:$] sampled_out_d1);									
\end{Verbatim}
\subsection{Soundness of abstraction}
We have taken a practical approach to establish that the abstraction used is \textit{safe} (does not miss real bugs) and \textit{appropriate} (does not find spurious bugs). The way we do this is by mutating the underlying design being verified to introduce a range of bugs. We then rerun the assertions in the model checker to check if the key properties (in this case the ordering property) fails. If it does, then the property used and the abstraction model is adequate and is safe. If the abstraction was too abstract, it would result in spurious failures on designs where there are no bugs. If the abstraction was too tight and could mask some of the bugs then during mutation the ordering property will not necessarily fail thereby flagging that the abstraction was not able to catch a real bug (which is inserted by mutation). We employ hand-mutation as the first port of call as we are able to identify a range of interesting control and data registers to mutate pretty easily --- something which is not yet possible with an automatic tool based mutation. Tool based mutations~\cite{certitude} are not clever enough to identify the control registers the way a human being can. Most tool-injected faults are quite simple stuck-at type faults, which are not adequate to unearth some potential bugs.

\subsection{Smart Tracker: Single Transaction Abstraction}
Using two transactions to model ordering checks is intuitive; however it is not necessarily the best possible way in terms of obtaining model checking performance. It needs two registers for modeling sampling in events and another two for sampling out events requiring in total 4 of these along with a read seen register. We now introduce what we call as a single transaction abstraction~\cite{DarbariDAC2014} designed using a counter. It is shown in Figure~\ref{two_abs_fig} on the left hand side. 

Our technique is based on tracking a watched data as it traverses inside the FIFO from the point it gets accepted on the input interface to the point it leaves the output interface of the design. The key idea in this abstraction however is to track \textit{only one} symbolic watched data (shown as $\mathtt{d}$ below) using a tracking counter. We do not store any other input data words (accepted in the design) in our testbench model. We only ever have to track the relative displacement of the trace of data words ahead of the watched data -- providing to us a very powerful method of abstracting away the storage (in our testbench) of all the other data words saving on state-space search for the model checker. 

Whenever a new data word gets accepted on the input interface the tracking counter (shown as $\mathtt{tracking\_counter}$ below)  increments by one if the data word is not the watched data. Whenever a stored data word in the FIFO design leaves it on a read, the tracking counter in the formal testbench decrements by one. When the watched data appears on the input interface and gets accepted inside the FIFO we freeze the increment of the tracking counter and only allow the counter to decrement on reads. This way at the time of sampling in the watched data we know how many data values were ahead of this watched data and therefore whenever the tracking counter reaches 1 we expect the watched data to appear at the output in the same or following clock cycle (depends on the implementation of the read logic in the FIFO design). If the watched data was to be ever reordered, lost, corrupted or duplicated - its history will be affected and the counter based check will be able to detect it thereby flagging the bug in the design. The model checker chooses all possible data words (at all possible time points) to be the watched data through symbolic processing of SAT based encoding of the combined design, our formal model, and any constraints and assertions. 

We now show the relevant aspects of the modeling code. As was the case with the two transaction abstraction, we keep the watched data arbitrary but fixed. This is done by using the $\mathtt{\$stable}$ operator in SVA. We also employ here an additional undriven wire $\mathtt{arbit\_window}$ to get extra non-determinism which gives us the ability to control which version of the watched data will be sampled in. 
The $\mathtt{sampled\_in}$ register is set when we have a write handshake ($\mathtt{w\_hsk}$), and the watched data has not been already sampled in (i.e. $\mathtt{incr}$ is high) and the watched data appears on the input port ($\mathtt{w\_data}$) and the undriven wire $\mathtt{arbit\_window}$ is set to 1 (i.e., this watched data is the data word that will be sampled in). It is quite possible that we have a write handshake and the very same watched data appears on the input port but the $\mathtt{arbit\_window}$ is not set to 1. In this case this instance of watched data will not be sampled in. The write and the read handshake signals $\mathtt{w\_hsk}$ and $\mathtt{r\_hsk}$ respectively control the increment ($\mathtt{incr}$) and the decrement ($\mathtt{decr}$) of the tracking counter. The $\mathtt{\$clog2}$ operator is a standard operator in SystemVerilog for calculating the logarithm to base 2. The parameter $\mathtt{DEPTH}$ is the depth of the FIFO whereas $\mathtt{WIDTH}$ is the size of the data word.  
\begin{Verbatim}[fontsize=\relsize{-2}, numbers=left, frame=single]
localparam DEPTH_BITS = $clog2(DEPTH);
logic [WIDTH-1:0] d;
assume property ($stable(d));
assign incr = w_hsk && !sampled_in;
assign decr = r_hsk && !sampled_out;
always @(posegde clk or negedge rst)
if (!rst)
  sampled_in <= 1'b0;
else if (w_data==d && incr && arbit_window)
  sampled_in <= 1'b1;
\end{Verbatim}
The register $\mathtt{sampled\_out}$ gets set when we are ready to sample out the watched data. This happens when the tracking counter is 1, and we have sampled in the watched data and we are ready to read ($\mathtt{r\_hsk}$ is high) and have not read out the watched data yet ($\mathtt{decr}$ is high). All of these conditions are defined by the signal $\mathtt{must\_read}$ shown below.
\begin{Verbatim}[fontsize=\relsize{-2}, numbers=left, frame=single]
assign must_read = (tracking_counter==1)  && sampled_in && decr;
always @(posegde clk or negedge rst)
if (!rst)
  sampled_out <= 1'b0;
else if (must_read)
  sampled_out <= 1'b1;
\end{Verbatim}
The tracking counter is then defined below. The counter will increment on a write handshake ($\mathtt{w\_hsk}$) provided we have not sampled in the watched data ($\mathtt{d}$) yet and would decrement on a read handshake provided we have not read out the watched data (i.e., $\mathtt{sampled\_out}$ register is set).
\begin{Verbatim}[fontsize=\relsize{-2},numbers=left,frame=single]
always @(posedge clk or negedge rst)
if (!rst)
  tracking_counter <= {DEPTH_BITS+1{1'b0}};
else
  tracking_counter <= tracking_counter + incr - decr;
\end{Verbatim}
The assertion to verify data integrity and ordering then becomes:\\
\begin{Verbatim}[fontsize=\relsize{-2}, frame=single, label=Master Assertion]
assert property (must_read |=> (r_data==d));
\end{Verbatim}
The assertion says that when the right condition for reading out the watched data is set then in the following clock cycle the watched data appears on the output data port. A one cycle delay in this case is due to a one cycle latency in the FIFO design around when read happens and when data comes out. 
If the bug in the design is related to data loss, data duplication, data corruption, or ordering the assertion will fail flagging an error with a counter example. This happens since the model checker tries to instantiate all possible data values to the watched data and chooses all time points to sample in the watched data. The liveness property (shown below) is also verified here. It checks that whatever data has gone in the FIFO comes out and is not lost. It does not however check for ordering correctness.\\
\begin{Verbatim}[fontsize=\relsize{-2}, frame=single, label=Liveness]
assert property (sampled_in |-> ##[0:$] sampled_out);									
\end{Verbatim}

\subsection{Invariants: Going beyond abstraction}
Both the above models are efficient when compared to methods where one uses another reference FIFO design as a testbench. However as the FIFO depths increase, despite the abstraction the performance of model checkers worsen. To cope with this, we devised a family of \textit{invariants} that we could prove as auxiliary helper asserts and when these are proven and later assumed it provides a dramatic turn around in proof times. The key invariants used are shown below. The invariants need information about the micro-architectural implementation of the FIFO which in our case happens to be the circular two pointer (read and write) based implementation and invariants once proven would be true of all reachable states from reset in our case. Invariants are also sometimes referred to as safety properties in the literature~\cite{DBLP:journals/toplas/ClarkeGL94} but when we refer to invariants we mean auxiliary properties that may be useful to prove the main safety or liveness property. In this way we do provide a dramatic increase in model checker performance as they help to reduce the state-space search.
\begin{enumerate}
\item [I4] Positional invariant tells the tool where  exactly in the DUT the tracked value is. This is calculated as function of read pointer and the tracking counter.
\item [I3] Once the tracked value is in the system then the tracking counter is in between the read and the write pointers.
\item [I2] If the tracking value has not entered the system then the counts between the DUT and the abstract model agree.
\item [I1] If the tracking value has not entered in the system then it couldn't have left it.
\end{enumerate}
The positional invariant is the key invariant required for proving the ordering property. As it turns out proving this key invariant is as hard if not harder sometimes than the ordering property itself. However, some other auxiliary properties are much easier to prove than the positional invariant and if these are proven and subsequently assumed then the proof times of the positional invariant scale linearly with the depth of the FIFO. We scripted an assume-guarantee flow which can take the easier invariants first to prove and then assume them to prove the harder ones. Using our flow, we will prove invariants in the following order (I4) followed by (I3) then (I2) and lastly (I1). What we additionally did was to use the same invariants with the two transaction abstraction model. We only applied this to one of the watched data words - the first to arrive in the design which in our case was $\mathtt{d1}$ and this was sufficient to give us scalability. Once the ordering property is proven we can assume it to prove the liveness property R4 which says all accepted input data words must eventually come out. 
\begin{figure}[!tp]
\begin{center}
\includegraphics[scale=0.37]{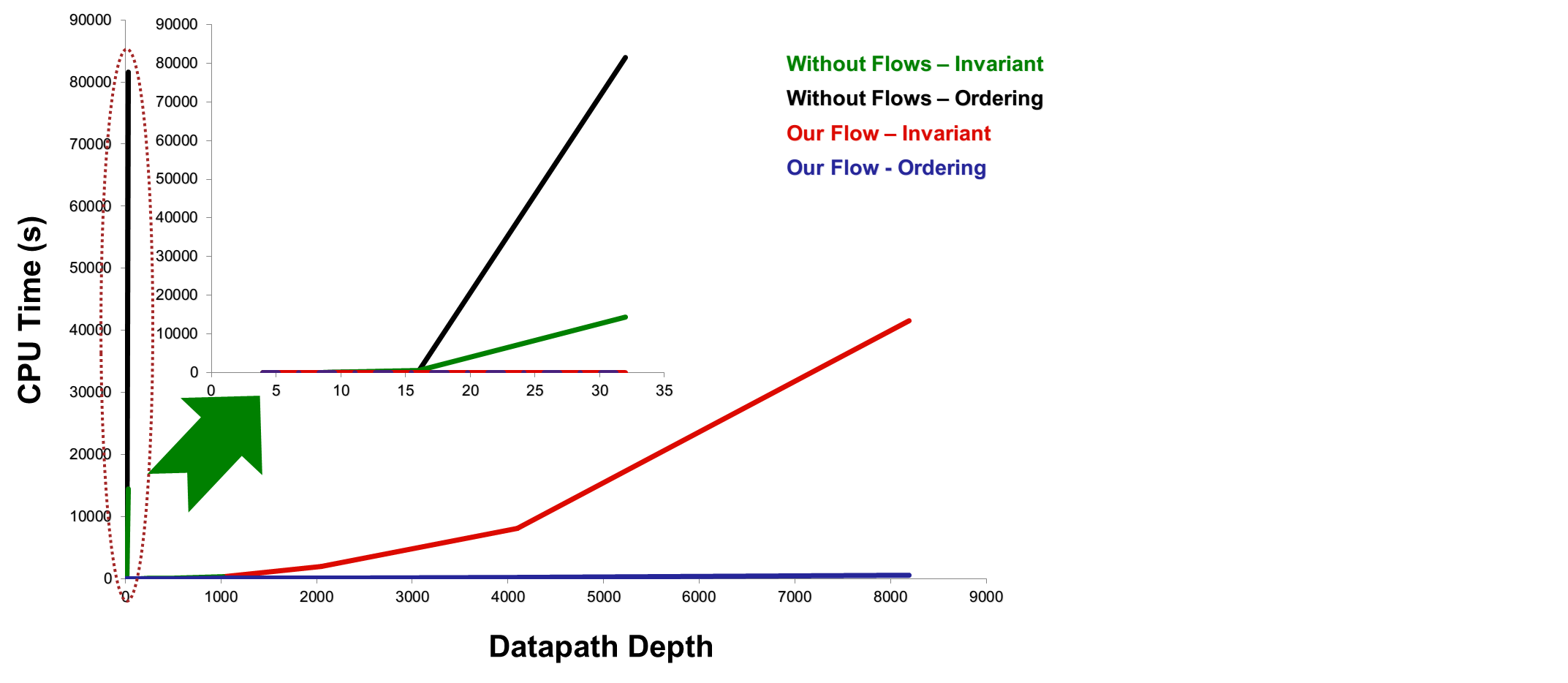}
\caption[justification=centering]{\scriptsize{This graph shows the time taken using the single transaction abstraction. On the Y-axis is the CPU time in seconds whilst on the X-axis is the DEPTH of the FIFOs. The curve shown in RED depicts the time taken to prove the positional invariant (I1) by proving and then assuming the other auxiliary invariants (I2), (I3) and (I4) first. Once I4 is proven the time it takes to prove the ordering property is shown in BLUE. The lines in black and green show what it takes to prove the positional invariant and the ordering property without using any assume-guarantee flow and auxiliary invariants. This results in the model checker running out of capacity after consuming 24 hours of CPU time for a 16 deep FIFO. The combined benefit of using abstraction, invariants and assume-guarantee is quite significant. We are able to verify ordering correctness property for FIFOs as deep as 8192 in 518 seconds. Moreover, using our approach the FIFO verification run time \textit{scales linearly} as the FIFO DEPTH \textit{scales exponentially} for the positional invariant. Most importantly for the key ordering property the curve for the run time remains nearly constant.}}
\label{fifograph} 
\end{center}
\end{figure}
\subsection{Results}
When we tested the smart tracker abstraction with assume-guarantee flow we obtained the results shown in Figure~\ref{fifograph}. This graph shows the time taken using the single transaction abstraction. On the Y-axis is the CPU time in seconds whilst on the X-axis is the DEPTH of the FIFOs. The curve shown in RED depicts the time taken to prove the positional invariant (I1) by proving and then assuming the other auxiliary invariants (I2), (I3) and (I4) first. Once I4 is proven the time it takes to prove the ordering property is shown in BLUE. The lines in black and green show what it takes to prove the positional invariant and the ordering property without using any assume-guarantee flow and auxiliary invariants. This results in the model checker running out of capacity after consuming 24 hours of CPU time for a 16 deep FIFO. The combination of using assume-guarantee flow together with invariants and abstractions allowed us to verify FIFOs as deep as 8192 the run time was 518 seconds (under 9 minutes!). 

\begin{figure}[!tp]
\begin{center}
\includegraphics[scale=0.6]{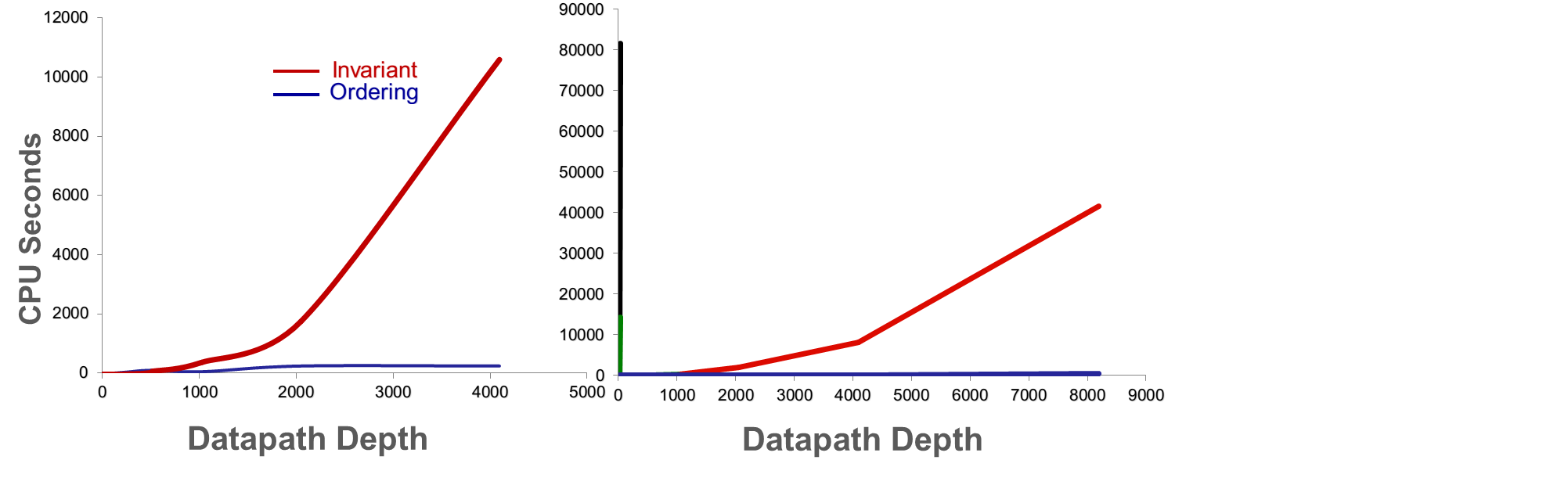}
\caption{\scriptsize{Comparison of single transaction and two transaction methods. Single transaction methodology (curve shown on right) is clearly more efficient than the two transaction (curve shown on left) as is obvious from the slope of the RED lines (which is less steep for single transaction) depicting proof times for the positional invariant in the graph.}}
\label{comparison-graphs}
\end{center}
\end{figure}

When we compare our two abstractions (using the assume-guarantee flow with both using invariants) we got the graphs shown in Figure~\ref{comparison-graphs}. As it is obvious from the graphs, the run time for the ordering property using both the abstractions is of the order of seconds. The combined benefit of using abstraction, invariants and assume-guarantee is quite significant and makes the FIFO verification 
run time \textit{scales linearly} as the FIFO DEPTH \textit{scales exponentially} for the positional invariant. Most importantly for the key ordering property the curve for the run time remains \textit{nearly constant} with increasing exponential DEPTH. The smart tracker abstraction is however more efficient than the two transaction abstraction as it has to perform checking with respect to only one symbolic watched data and not two. This creates much less model checking overhead for smart tracker abstraction. By proving and then assuming the ordering property we are able to prove the liveness property within seconds/minutes of wall-clock time.

We found several bugs in different FIFOs under development. Most of these were functional bugs but notably the bug that caught the most attention with designers was a bug related to redundant area (consuming dynamic power) in a low power optimized FIFO design.
\section{A Simple Arbiter}\label{simplearb}
\begin{figure}[!tp]
\begin{center}
\includegraphics[scale=0.25]{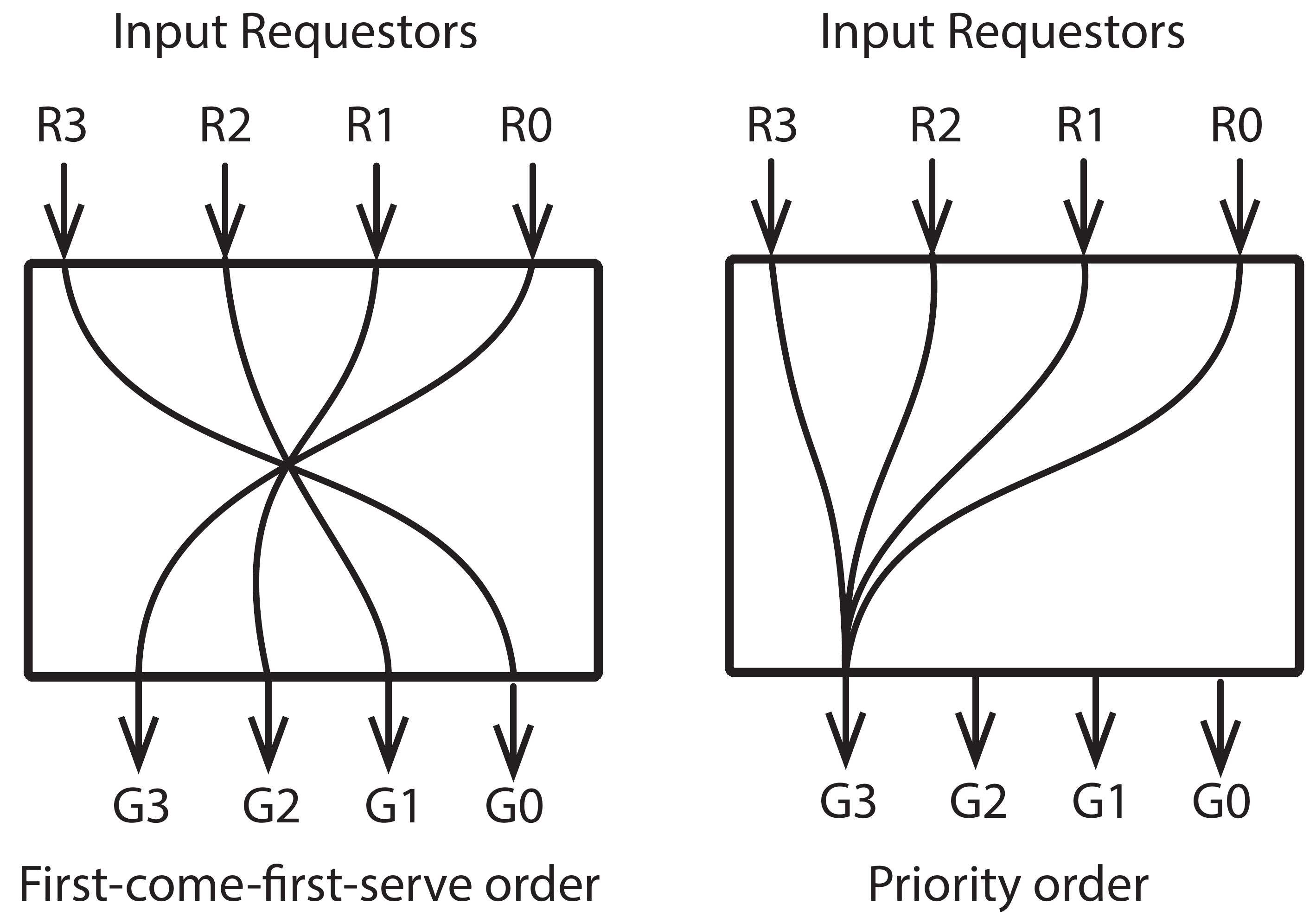}
\caption{\scriptsize{A Simple Arbiter. It has interesting arbitration rules to check and even though does not have a lot of state in it; still has stallers, FIFOs and FSMs. To verify a design of this size with high confidence using dynamic simulation requires non-trivial investment of effort, is not necessarily reusable, and will certainly take a lot longer to converge in terms of getting good coverage than the results we obtained using our formal modeling.}}
\label{simple_arb}
\end{center}
\end{figure}
We now present a case study of a simple arbiter which arbitrates on four incoming requestors. This is shown in Figure~\ref{simple_arb} by four input interfaces R0 to R3. Requests are accepted through a handshake protocol and are directed to any of the four outputs G0 to G3 where they get granted again on a handshake. On the input side, in a clock cycle one can have up to 4 handshakes, meaning all 4 input requestors may wake up in a single clock cycle but on the output side, at most one handshake is allowed in a single clock cycle. It is not a requirement that on every clock cycle we must have input/output activity. It is quite possible that once incoming requests are accepted they may not be granted for several clock cycles on the output interface.

In Figure~\ref{simple_arb} on the left is the case where a stream of transactions originate from input requestors going towards different outputs. It does not matter if one input requestor got accepted before the other. However, the rule is that whatever got accepted at each input interface must be transferred to the correct output interface and in {\it the same order in which they were accepted at the input interface}. On the right side of the figure  we show the case when all of the input requestors are competing for the same	output G3 and all of the input requests were accepted in {\it the same clock cycle.} In this case a priority order is enforced and the requestor with a lower index has a higher priority. Thus in the diagram shown on the right, R0 will be granted access to G3 before R1, R1 granted before R2, and R2 granted before R3. If however requests originated on two different input requestors in {\it two different clock cycles} being sent to the {\it same output interface} then first-come-first-serve has precedence over priority ordering. For example if R1 was accepted before R0 and they were both competing for G3 then R1 would be granted access to G3 before R0. The last rule which is not very difficult to verify is the case when different input requestors send requests to different output interfaces. In this case it is just sufficient to check that these requests do get granted, but this check need not be encoded explicitly as the first-come-first-serve check that we model checks that no data is dropped, duplicated or corrupted and is correctly ordered and that the arbitration is fair.

Our approach in verifying this arbiter is similar to verifying a FIFO in that we will use the concept of a watched value to observe and track. However, whereas in the FIFO we were watching a data value, in this case we will observe a symbolic watched requestor ($\mathtt{watched\_req}$). The other key signals used in our model are the usual suspects here - sampling registers ($\mathtt{sampled\_in}$ and $\mathtt{sampled\_out}$), wires ($\mathtt{incr}$ and $\mathtt{decr}$) for increment and decrement respectively, $\mathtt{arbit\_window}$ for creating non-determinism and a new wire ($\mathtt{leading\_req}$) that calculates how many requests are leading (ahead) of the watched requestor. Below we declare a watched requestor ($\mathtt{watched\_req}$) and declare it to be stable.
\begin{Verbatim}[fontsize=\relsize{-2}, numbers=left, frame=single]
logic                      watched_req;
assume property ($stable(watched_req));
\end{Verbatim}

On every incoming request into the arbiter on a non-watched port increment the counter. The handshake signals $\mathtt{hsk\_in}$ and $\mathtt{hsk\_out}$ get high if one or more handshakes happen on the input or output respectively. Once a handshake happens on the watched requestor freeze the counterӳ increment. If multiple requests arrive at the same time as the requests on the watched requestor,  a mask is used to only increment the counter by the number of requestors with greater than or equal to priority as the watched requestor. This then keeps track of where the request from the watched requestor sits in the arbitration queue. You only ever need to monitor the first request on the watched requestor. On every handshake at the output port for the non-watched requestor decrement the counter. When the counter becomes one expect the data from the watched requestor to appear at the output port. The counter's increment and decrement needs adjustment based on priority encoding. This in our example becomes the peripheral code which can be tweaked. The SVA built-in function $\mathtt{\$countones}$ simply counts the number if 1s in a given vector.
\begin{Verbatim}[fontsize=\relsize{-2}, numbers=left, frame=single]
assign incr             = sampled_in  ? 'h0 : leading_req;                    
assign leading_req      = ready_to_sample ? ($countones(hsk_in & mask[watched_req])):$countones(hsk_in);				
assign decr             = |hsk_out  && !sampled_out;
assign ready_to_sample  = hsk_in[watched_req] && arbit_window;
always @(posedge clk or negedge rst)
if (!rst)
  sampled_in <= 1'b0;
else if (ready_to_sample)
  sampled_in <= 1'b1;
always @(posedge clk or negedge rst)
if (!rst)
  sampled_out <= 1'b0;
else if (sampled_in && hsk_out[watched_req])
  sampled_out  <= 1'b1;
\end{Verbatim}
With the increment and decrement now adjusted for this design accordingly the tracking counter now looks similar to the counter in the smart tracker abstraction. The term that decides when it is permissible to observe the watched requestor's data out is given by the term $\mathtt{must\_read}$.
\begin{Verbatim}[fontsize=\relsize{-2}, numbers=left, frame=single]
always @(posedge clk or negedge rst)
if (!rst)
  tracking_counter <= 'h0;
else 
  tracking_counter <= tracking_counter + incr - decr;
assign must_read = (tracking_counter==1) && sampled_in && hsk_out_glbl;
\end{Verbatim}
The only check needed to verify all the required arbitration requirements is shown below:\\
\begin{Verbatim}[fontsize=\relsize{-2}, frame=single, label=Master Assertion]
assert property (sampled_in && hsk_out_glbl && (tracking_counter == 1) |=> sampled_out);
\end{Verbatim}
The scoreboard  and a single assertion is able to check all the key verification requirements as follows:
\begin{enumerate}
\item Starvation: By using a symbolic watched requestor, the solution will check all requestors and make each one the watched requestor. If any requestor was starved access then it would never be seen coming out of the arbiter and we would see the assertion failing.
\item Fairness: The solution encapsulates the method to check for fairness automatically. This is done by checking that each requestor is serviced when it is expected to be. This means that if any requestor was serviced unfairly with respect to the arbitration scheme then the assertion would fail.
\item Data-Integrity: Is checked by ensuring that the correct data comes out at the output when that requestors data is expected.
\item Race freedom: Because this is based around enforcing the priority and checking every requestor is serviced when it should be, if there were any race conditions which resulted in incorrect order of service then the assertion would fail.
\item Deadlock freedom: If the system was to get into deadlock then it would not be able to process any more requestors and we would not see the watched requestor on the output port in some scenarios. Because by using formal all scenarios are tested there would be an assertion failure in the case of deadlock.
\end{enumerate}
\subsection{Results}
We tested our solution on a range of arbiters including priority based, first-come-first-serve and round-robin. In all of these cases our solution was able to obtain exhaustive proofs of the main assertion within seconds of wall-clock time for up to 4-8 requestors. When we increased the requestors to up to 128, the run times for exhaustive proofs were of the order of minutes. We also found bugs in some of these arbiters.

\section{A Memory sub-system arbiter}\label{mem_arbiter}
\begin{figure}[!tp]
\begin{center}
\includegraphics[scale=0.80]{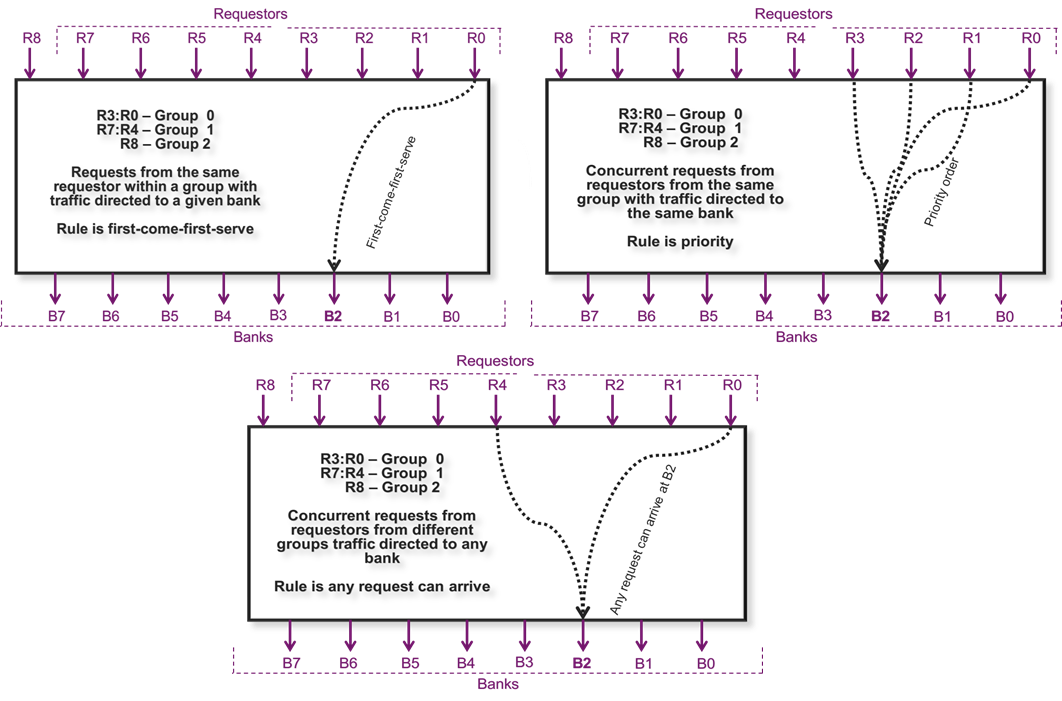}
\caption{\scriptsize{A Memory sub-system arbiter's high level block diagram showing at an abstract level how multiple concurrent transactions originate from the requestors R0 to R8 directed towards the memory banks B0 to B7.}}
\label{mem_sys}
\end{center}
\end{figure}
We now show how we apply the arbiter and FIFO verification methodology for verifying a complex memory sub-system arbiter. Arbiters such as this one are complex as they have plenty of control and state needed to achieve high-performance arbitration. Typical data structures used in the design include register files, FIFOs, random stallers, and FSMs. The mesh of all of these creates a significant verification challenge for both simulation and formal. Whereas with formal we are able to get a rigorous proof of correctness it is not without a good methodology. Figure~\ref{mem_sys} shows at an abstract level how multiple concurrent transactions originate from the requestors R0 to R8 directed towards the memory banks B0 to B7. This arbiter has three groups of incoming requestors, arbitrating over 4 groups of memory banks organized as 2 banks in each group. For simplicity we flatten the incoming requestors and label them R0 to R8 and the memory banks as B0 to B7. In practice, the requestors R0 to R3 are in one group; R4 to R7 in another whilst R8 is the last group. Each requestor within R0 to R3 or R4 to R7 is also referred to as a requestor instance. 

The key arbitration rules~\cite{DarbariJUG} that are challenging to verify for this arbiter are:
\begin{enumerate}
\item First-come-first-serve: Requests originating from any requestor instance from any group sent to any of the memory banks must arrive at the bank in the same order as they were received at the input requestor.
\item  Priority order: Requests originating from the requestor instances of the same group competing for the same memory bank must have a priority order. For example if multiple instances from the group R0 to R3 sent requests in the same cycle to a bank $B_i$ where $i$ ranges from 0 to 7; then requests from R0 should be accepted before requests from R1, requests from R1 before R2, and R2 before R3. The same rule applies to requestor instances from the second requestor group R4 to R7.The last group has a lone instance so priority order rule is not applicable to it.
\item Out-of-order: If requests from different requestor groups competed for the same bank then any of the accepted requests can arrive at the given bank in any order. 
\end{enumerate}
\subsection{Scenario splitting abstraction}
We exploit a new technique here to isolate the different scenarios. The first scenario can focus only on the first-come-first-serve behaviour and mask away the tracking or observation of any multi-instance requests. The second scenario can focus on multi-instance activity but narrow down the observation to only those requests that are directed to the same bank and mask away any tracking and observation logic for other scenarios. The last scenario would be to observe traffic originating from any requestor going to any  bank and ensuring that these requests are received and not lost. However, this scenario is automatically subsumed by the previous two scenarios which are a lot more stricter where we can prove that both first-come-first-serve and priority ordering is correct. If these behaviours are true then it is certain that all requests are arriving at all the banks.

The use of scenario splitting is another form of abstraction in that it allows one to focus the verification to be on one scenario at a time and allows us to abstract away the non-interesting scenarios. This is the key to our approach in verification of these designs where a combined tracking and observation logic encapsulated in a single model (as in the previous section) is limiting for performance and tractability of the formal proofs. Once we set out to build the formal testbench by exploiting scenario splitting the rest of the methodology falls in place directly from the previous sections - combination of the smart tracker abstraction for first-come-first-serve verification and the two transaction abstraction and its extensions for proving priority order. The basic data structures we use in the modelling of the assertions are shown below. It is obvious looking at these data structures that the general scheme here is to replicate the FIFO single transaction method using in this case an array of tracking counters, in total requiring 64 $\mathtt{((REQ-1)\times(INST-1)\times(GRP-1)\times(BNK-1))}$ tracking counters each $\mathtt{MAX+1}$ wide to count $\mathtt{MAX}$ transactions. 
\begin{Verbatim}[fontsize=\relsize{-2}, numbers=left, frame=single]
wire [BNK-1:0]  ready_to_sample_out [REQ-2:0][INST-1:0][GRP-1:0];
wire [3:0]      decr_tc             [REQ-2:0][INST-1:0][GRP-1:0][BNK-1:0];
wire [INST-1:0] hsk_in              [REQ-2:0];
wire [BNK-1:0]  hsk_out             [GRP-1:0];
reg  [MAX:0]    tracking_cnt        [REQ-2:0][INST-1:0][GRP-1:0][BNK-1:0];
reg  [BNK-1:0]  seen_in_watched_id  [REQ-2:0][INST-1:0][GRP-1:0];
reg  [BNK-1:0]  seen_out_watched_id [REQ-2:0][INST-1:0][GRP-1:0];
\end{Verbatim}
Just as we had ready to sample in and ready to sample out registers in FIFO verification, we have these here as well. The only difference here is that we have an array of these just as we have an array of increment, decrement and sampled in and sampled out registers. The basic principle remains the same in that the ready to sample out registers get set to 1 when the corresponding tracking counter ($\mathtt{tracking\_{cnt}}$) is reduced to 1 and we have sampled in the watched identifiers ($\mathtt{seen\_in\_watched\_id}$) and we are decrementing (meaning we haven't seen the watched id appear at the expected bank and we are having a handshake at the bank interface akin to reading the watched id out akin through a FIFO pop).

\noindent The code shown below keeps track of what ids have been seen in at the input interface directed to which group and bank. This is done by employing a $\mathtt{generate-for}$ loop which tracks where the requests are going to (indexed by group ($\mathtt{g\_i}$) and bank ($\mathtt{b\_i}$)) and originating from a given requestor and instance which is shown for requestor index 0 and instance 0.
\begin{Verbatim}[fontsize=\relsize{-2}, numbers=left, frame=single]
generate
 for (g_i=0; g_i<GRP; g_i=g_i+1)
  for (b_i=0; b_i<BNK; b_i=b_i+1)
  assign hsk_out [g_i][b_i]  = group_bank_valid[g_i][b_i] && group_bank_enable[g_i][b_i];
always @(posedge clk or negedge rst)
if (!rst)
  seen_in_watched_id[0][0][g_i][b_i] <= 1'b0;
else if (hsk_in[0][0] && req_out[0][0][g_i][b_i] && (data[0]==watched_id) && !other_req_active[0])
  seen_in_watched_id[0][0][g_i][b_i] <= 1'b1;
endgenerate
\end{Verbatim}											
We now show the array of wires for increment and decrement indexed per requestor, per instance, per group, and per bank. This ensures we keep track of all requests originating from each requestor on an instance basis and track which group and bank they are directed to. This needs to be done for the sample out registers ($\mathtt{seen\_out\_watched\_id}$), $\mathtt{ready\_to\_sample\_out}$ wires as well as the tracking counters ($\mathtt{tracking\_cnt}$).
\begin{Verbatim}[fontsize=\relsize{-2}, numbers=left, frame=single, baselinestretch=1]
generate
 for (r_i=0; r_i<REQ-1; r_i=r_i+1)
  for (inst_i=0; inst_i<INST; inst_i=inst_i+1)
  assign hsk_in[r_i][inst_i] = req_valid[r_i][inst_i] && req_enable[r_i][inst_i];
    for (g_i=0; g_i<GRP; g_i=g_i+1)
     for (b_i=0; b_i<BNK; b_i=b_i+1)
     assign incr_tc[r_i][inst_i][g_i][b_i] = hsk_in[r_i][inst_i] && req_out[r_i][inst_i][g_i][b_i];
     assign decr_tc[r_i][inst_i][g_i][b_i] = !seen_out_watched_id[r_i][inst_i][g_i][b_i] &&
                                              hsk_out[g_i][b_i];
     assign ready_to_sample_out[r_i][inst_i][g_i][b_i]  = (tracking_cnt[r_i][inst_i][g_i][b_i]=='h1) &&
                                                           seen_in_watched_id[r_i][inst_i][g_i][b_i] &&
		                                                       decr_tc[r_i][inst_i][g_i][b_i];
always @(posedge clk or negedge rst)
 if (!rst)
  seen_out_watched_id[r_i][inst_i][g_i][b_i] <= 1'b0;
 else if (ready_to_sample_out[r_i][inst_i][g_i][b_i])
  seen_out_watched_id[r_i][inst_i][g_i][b_i] <= 1'b1;

always @(posedge clk or negedge rst)
 if (!rst)
  tracking_cnt[r_i][inst_i][g_i][b_i] <= {MAX+1{1'b0}};
 else
  tracking_cnt[r_i][inst_i][g_i][b_i] <= tracking_cnt[r_i][inst_i][g_i][b_i] +
                                         incr_tc     [r_i][inst_i][g_i][b_i] -
                                         decr_tc     [r_i][inst_i][g_i][b_i];
endgenerate
\end{Verbatim}
We now show the template assertion for checking first-come-first-serve property. Using a generate loop 64 of these are generated - from the cross product of the set of multi-instance requestors ($\mathtt{REQ-1}$) and number of groups  and banks ($\mathtt{(GRP-1) \times (BNK-1)}$).\\
\begin{Verbatim}[fontsize=\relsize{-2}, numbers=left, frame=single, label=Master assertion for Ordering]
generate
 for (g_i=0; g_i<GRP; g_i=g_i+1)
  for (b_i=0; b_i<BNK; b_i=b_i+1)
    assert property (ready_to_sample_out[0][0][g_i][b_i] && !other_req_active[0] 
		                 |->
		    (OUT_ID[g_i][b_i] == watched_id));
endgenerate
\end{Verbatim}
 The sample assertion shown here is one of the 8 ($\mathtt{(REQ-1) \times INST}$) assertions coded explicitly in the generate loop. The $\mathtt{for}$ loops make sure that each of the explicitly coded assert within the generate loop is unrolled for each bank ($\mathtt{BNK}$) for all groups ($\mathtt{GRP}$). We maintain a side condition in the antecedent of the assertions using a predicate $\mathtt{!other\_req\_active}$ which ensures that only one fixed instance of a fixed requestor is actively sending the requests and not any other instance in a given clock cycle. This is to ensure we mask away any noise that is not needed in proving the property in question which is that a stream of requests originating from any requestor instance remains ordered on arrival at a given bank.

By keeping this guard we are able to seperate out different scenarios for first-come-first-serve checks for each bank with respect to each requestor instance. There are 8 such predicates 
($\mathtt{(REQ-1) \times INST}$), where $\mathtt{REQ}$ is fixed at 3 and $\mathtt{INST}$ at 4 for this design. We would like to point out that for the requestor with a single instance we don't have to harness any generate loop hence we have $\mathtt{REQ-1}$. Each of these assertions prove within an hour of wall clock time.

We now show how we structure the properties that reason about multi-instance activity from a requestor group. As pointed out there are two requestor groups with 4 instances each and a third requestor group with a lone instance. The challenge is in proving properties which check that ``if multiple instances in the same requestor group accepted incoming data being sent to the same bank then the priority order on instances is respected with respect to the bank". We decompose this overall check into 11 different cases 
We now show how we structure the properties that reason about multi-instance activity from a requestor group. As pointed out there are two requestor groups with 4 instances each and a third requestor group with a lone instance. The challenge is in proving properties which check that ``if multiple instances in the same requestor group accepted incoming data being sent to the same bank then the priority order on instances is respected with respect to the bank". We decompose this overall check into 11 different cases 
($\mathtt{CASES}$=11) per requestor group ($\mathtt{REQ}$=2) with multiple instances (in our case these are 2) per group ($\mathtt{GRP}$=4) per bank ($\mathtt{BNK}$=2). In total we have 
$\mathtt{REQ \times GRP \times BNK \times CASES}$ = 176 assertions generated from generate loops coded in the following manner.  The number 0 in 
$\mathtt{seen\_multi\_inst [r\_i][g\_i][b\_i][0]}$ denotes the case split number; it is the initial case that talks about instances 0 and 1 of a requestor $\mathtt{r\_i}$ sending data to a bank $\mathtt{b\_i}$ in the same clock cycle. We have six such cases of two instances active at the same time. We have four cases of three instances active at a time, and one case of all four instances sending data to the same bank. \\
\begin{Verbatim}[fontsize=\relsize{-2}, numbers=left, frame=single, label=Master assertion for Priority]
reg [CASES-1:0] seen_multi_inst[REQ-1:0][GRP-1:0][BNK-1:0];
generate
  for (r_i=0;r_i<REQ-1;r_i=r_i+1)
   for (g_i=0;g_i<GRP;grp_i=g_i+1)
    for (b_i=0;b_i<BNK;b_i=b_i+1)
as_check_arbitration_i0_and_i1:
assert property (seen_multi_inst [r_i][g_i][b_i][0] && !req_out [r_i][0][g_i][b_i] 
                 |->
	         !req_out [r_i][1][g_i][b_i]);
endgenerate
\end{Verbatim}
Each assertion comes with almost a dual-like cover property which helps to sanity check for those scenarios where the proof of the above assertion may come out as a valid proof but only due to vacuity. Vacuity can happen due to the way implication properties work in SVA specifically and logic in general; where a failing antecedent in an implication can cause the overall proof to be declared valid. To rule those cases out we just write an adjoining cover to ensure that it is valid and does not fail.
\subsection{Results}
It should be noted how the above property is identical to the two transaction based FIFO verification method we showed. All of the generated properties prove within 5-7 minutes of wall-clock time. In practice we just select all of these at once and within 5-7 minutes we are able to finish off all the 176 proofs. Similarly for the 64 ordering properties we prove all of these at once and within an hour of wall-clock time we are done with the exhaustive proofs.

We would like to emphasize that we applied our methodology on this design once extensive simulation had been done and bugs had been fixed. We did not find any further bugs in this design; we did however exhaustively prove that there were no bugs in implementation of the key arbitration rules outlined above which made the whole methodology very powerful as it provided the much needed assurance levels.
\section{Synchronizer}\label{sync}
In this section we will show how the smart tracker abstraction of Section~\ref{fifo} is directly applicable in verifying synchronizers. A synchronizer is a device used in any environment where multiple clocks are used, specially when these clocks may have edges quite close to each other. This gives rise to a multitude of problems such as meta-stability, reconvergence, and loss of data. A commonly used design implementation for a synchronizer uses a gray-coded FIFO driven from multiple clocks from different domains. There are a range of checks that are exercised to validate these designs such as read/acknowledge behaviour on both write and read interfaces, metastability checks and checks on data integrity. Verification of data-integrity, ordering and proving absence of data loss is again a key requirement that not only requires exhaustive proofs and high assurance, it is also a challenge for both simulation and formal due to multiple clocks and possible combination of meta-stable events causing problems with data.  

We show the important bits of the modelling code to highlight our solution. The important thing to note below is the use of different clocks - write clock ($\mathtt{wclk}$) and read clock ($\mathtt{rclk}$) in modelling the sampling in and sampling out events. Other than that, the basic declarations of wires such as $\mathtt{arbit\_window}$, $\mathtt{incr}$, $\mathtt{decr}$, and $\mathtt{must\_read}$ remain here just like in the smart tracker abstraction. We also keep the watched data word stable and assume the existence of write and read handshake signals driven by request and acknowledge pairs on the input and output interfaces respectively. The registers used for defining the start and the stop events are the usual sampled in and sampled out registers. 
\begin{Verbatim}[fontsize=\relsize{-2}, numbers=left, frame=single]
localparam DEPTH_BITS = $clog2(DEPTH);
logic [WIDTH-1:0] d;
assume property ($stable(d));
assign incr = w_hsk && !sampled_in;
assign decr = r_hsk && !sampled_out;
always @(posegde wclk or negedge rst)
if (!rst)
  sampled_in <= 1'b0;
else if (w_data==d && incr && arbit_window)
  sampled_in <= 1'b1;
always @(posegde rclk or negedge rst)
if (!rst)
  sampled_out <= 1'b0;
else if (r_data==d && decr)
  sampled_out <= 1'b1;
\end{Verbatim}
We now show how the tracking counter is implemented in this case by employing two different counters - one that tracks the outstanding write count ahead of the watched data word ($\mathtt{w\_count}$) and other that tracks the outstanding read count ($\mathtt{r\_count}$). The tracking counter itself is simply the difference between ($\mathtt{w\_count}$) and ($\mathtt{r\_count}$). 
\begin{Verbatim}[fontsize=\relsize{-2}, numbers=left, frame=single]
always @(posedge wclk or negedge rst)
if (!rst) 
  w_count <= {DEPTH_BITS+1{1'b0}};
else
  w_count <= w_count + incr;
always @(posedge rclk or negedge rst)
if (!rst) 
  r_count <= {DEPTH_BITS+1{1'b0}};
else
  r_count <= r_count + decr;
assign tracking_counter = w_count - r_count;
\end{Verbatim}

The assertion to verify data integrity and ordering relies on using $\mathtt{must\_read}$ which gets the value 1 when the tracking counter has reduced to 1 and the watched data word was sampled in before and there was a read handshake in the current clock cycle of the read clock.
\begin{Verbatim}[fontsize=\relsize{-2},frame=single]
assign must_read = tracking_counter==1  && sampled_in && decr;
\end{Verbatim}

The assertion then is shown below. It is worth pointing out that this is explicitly clocked by the positive edge of the read clock $\mathtt{rclk}$. \\
\begin{Verbatim}[fontsize=\relsize{-2},  frame=single, label=Master Assertion]
assert property (@(posedge rclk) must_read |=> (r_data==d))
\end{Verbatim}

\subsection{Results}
We tested our solution on several clock ratios for varying depths of the synchronizer FIFO and the results were similar to what we showed for the FIFO using smart tracker abstraction. There was however a delta increase in time over the FIFO verification runtimes due to multiple clocks being used here and we did find this delta increase linearly as clock ratios increased from 1:2 to 1:32.
\section{A Packet Based Design}\label{packet}
In this section we will introduce our last case study of this paper. This design is also a data transport design but unlike a FIFO or an arbiter where the data does not change in size as it travels from input to the output; in this design the data can be \textit{dropped, shrunk or grown}. This design is representative of a family of designs used in networking domains where data packets regularly go through this form of elasticity.

The drop capability allows the design at any time to drop the packet at the head of the queue. The elastic capability to shrink or grow the data is controlled by an input called $\mathtt{elasticate}$. To decide when to resize the data we check if the packet queue is greater than {\it half full}. If it is half full and the input $\mathtt{elasticate}$ is high then the design looks at the lower three bits of the packet at the head of the queue. Having decoded this the output mask will be updated and the number of valid data bits read out in the future will change accordingly for all the packets in flight.

The modelling code shown below shows the key fragments of our abstract model --- note the addition of an internal mask signal ($\mathtt{internal\_mask}$) that captures the snapshot of how much shrinking/growth is to be done when appropriate conditions are set on the input. The internal mask is used to decide what fraction of a full packet size is being read out. The bits are reset to all 1 so the entire (100\%) packet size should be checked. If the update packet size conditions are met then the data at the head of the queue is checked. If the value of these bits is greater than or equal to the current value of $i$ then that bit will be set to 1 otherwise it will be set to 0. So if the head of the packet was of value 2 then the internal mask becomes $\mathtt{00111}$. This means we now expect the packet to be 60\% of its original size. 
\begin{Verbatim}[fontsize=\relsize{-2}, numbers=left, frame=single]
reg [SHRINK_VECTOR-1:0] internal_mask;
genvar i;
  generate for (i=0;i<SHRINK_VECTOR;i++)
    always @(posedge clk or negedge rst)
      if (!rst)
        internal_mask[i] <= 1'b1;
      else if ((in_flight_counter > ((2**DEPTH_BITS)/2)) && elasticate)
           if ({data[rptr][INTERNAL_MASK_HEADER-1:0]} >= i)
             internal_mask[i] <= 1'b1;
           else internal_mask[i] <= 1'b0;
  endgenerate
\end{Verbatim}
This value of internal mask gets transformed into an actual mask that can be applied directly to the data at the output. This is defined by $\mathtt{output\_mask}$ shown below. A counter $\mathtt{in\_flight\_counter}$ is used in the design to just keep track of how many packets are there in the queue. It increments on every write and decrements on a read or a drop. Please note that this counter in the design should not be confused with the tracking counter used in our abstract model of the formal test bench which counts how many values are ahead of the watched data.
\begin{Verbatim}[fontsize=\relsize{-2}, numbers=left, frame=single]	
wire  [DATA_WIDTH-1:0]         output_mask;
wire  [DATA_WIDTH-1:0]         output_mask_stretched;
assign mask_o                = internal_mask;
assign output_mask_stretched = (2**(mask_o+1))-1;          
assign output_mask           = (2**(internal_mask+1))-1; 
\end{Verbatim}
The rest of the abstract model now is very similar to how we modelled it for the FIFO with a slight modification to the model to take into account the role of the $\mathtt{drop}$ signal which can drop the data. This affects the model shown below in calculating assignments to $\mathtt{decr}$ and the $\mathtt{ready\_to\_sample\_out}$ signals. The other signals used in the model below are the write handshake ($\mathtt{w\_hsk}$), the read handshake ($\mathtt{r\_hsk}$) and the sampling registers -- $\mathtt{sampled\_in}$ and $\mathtt{sampled\_out}$ -- all defined in the usual way.
\begin{Verbatim}[fontsize=\relsize{-2}, numbers=left, frame=single]
assign incr = w_hsk && !sampled_in;
assign decr = (r_hsk || drop) && !sampled_out;
assign ready_to_sample_out   = (tracking_counter == 1) && sampled_in && decr && !drop;
\end{Verbatim}
The pivotal tracking counter used in the formal test bench is then defined as shown below.
\begin{Verbatim}[fontsize=\relsize{-2}, numbers=left, frame=single]
always @(posedge clk or negedge rst)
 if (!rst)
  tracking_counter <= {(DEPTH_BITS+1){1'b0}};
 else tracking_counter <= tracking_counter + incr - decr;
\end{Verbatim}
The assertion shown below establishes that every packet of data that has been read into the design and has not been dropped is correctly read out, is not duplicated, is of the correct size, and remains correctly ordered. Note that the check takes into account the packet resizing by performing a bitwise AND with the appropriate mask registers.\\

\begin{Verbatim}[fontsize=\relsize{-2}, frame=single, label=Master assertion]
assert property (ready_to_sample_out |=> ((data_o & output_mask_stretched) == (d & output_mask)));
\end{Verbatim}
\subsection{Results}
\begin{figure}
\begin{center}
\includegraphics[scale=1.0]{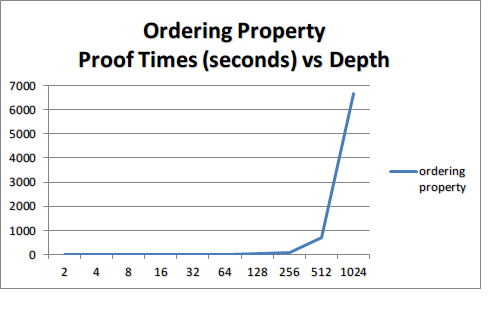}
\caption{Results showing time taken to prove the master assertion vs depth of the pipe using assume guarantee flow and invariants for a 32-bit wide data packet. Time here is not CPU time but real time in seconds. We should point out that in case of FIFO the proof times for its master assertion remained nearly constant (see Figure~\ref{fifograph}); however in case of this design though the time taken is still of the order of seconds, it does \textit{scale linearly} and the slope of this increases linearly with \textit{increasing exponential depth} of the pipe. This is due to the fact that packets in this design may go through an elastic shrink-grow phase which increases the complexity and the number of combinations to check.}
\label{packet_results}
\end{center}
\end{figure}
Using the same invariant flow as described in Section~\ref{fifo}, we have been able to exhaustively prove these features on this design up to 1024-deep. The time consumed in proving the master assertion is of the order of seconds of real time, shown in Figure~\ref{packet_results}. This is possible by using abstraction, invariants and our assume-guarantee flow. The class of invariants used here is similar to the ones we used for FIFO verification. We would like to point out here that we carried out testing for a fixed depth of the pipe against variable data width. That curve is also linear in time. Due to lack of space we cannot show this here.
\section{Related Work}\label{related}
In this section we present some comparison points with respect to two other closely related pieces of work. We would like to outline the key similarities and differences with these and would like to point out how our method is new. We will provide these comparisons with respect to the following aspects: (a) Use of symbolic variables (b) Formal models using counters (c) Invariants and Assume-Guarantee reasoning and (d) Scenario splitting.
\subsection{Use of symbolic variables}
The use of symbolic variables to watch non-deterministically values or ports is similar to how it has been done in the paper~\cite{DBLP:conf/fmcad/AggarwalCKS11}. However the way we extended this by adding a counter based abstraction is new. Symbolic variables are not used in this way by Benalycherif~\cite{DBLP:journals/entcs/BenalycherifM09}. The impact of our work using counters to perform abstraction and adding states to our model becomes more significant in the light of the speed up we obtained with our smart tracker model for FIFO verification. In~\cite{DarbariCDNLive2014} we reported a speed up of 500 times compared to the best known commercial solution. Considering that conventional literature on model checking and abstraction often considers counters to be bad for abstraction and as a consequence they are black-boxed (meaning not complied in the design) this is a significant new finding. This is also true of designs with FIFOs where conventional wisdom is about blackboxing them away. Our view is that in practice when these components are abstracted away not only do we end up chasing spurious bugs but also it is not practically sensible to verify designs without them when in fact a lot of bugs could be because of them in the first place.
\subsection{Formal models using counters}
Our approach is similar to the one proposed by \cite{DBLP:journals/entcs/BenalycherifM09} in that we both choose to use counting as pivotal way of modelling the main property for checking ordering although \cite{DBLP:journals/entcs/BenalycherifM09} never really modelled explicitly any counters. In fact in their approach the authors have not used modelling at all. In \cite{DBLP:journals/entcs/BenalycherifM09} the authors have formulated the theoretical connection between data independence and verification of ordering based components. They identified two key conditions which they call sufficient conditions for identifying data independence and also showed how a design such as a FIFO can be verified. The authors argue that checking that ``output data streams are samplings of input data streams" cannot be directly checked in a practicable manner using the constructs of temporal property languages such as PSL and SVA and argue that modelling code is required. This is something we agree with; however we do not agree with the authors' point of view that using glue logic in some modelling language is a bad thing. In fact we found that a lot of scalability in formal comes precisely by way of identifying how to model things more efficiently with glue logic and also allows one to have better controllability and observability something that is also encouraged by \cite{DBLP:conf/fmcad/AggarwalCKS11}.  We were able to scale our method massively to very deep FIFOs and we are not sure how the method proposed by \cite{DBLP:journals/entcs/BenalycherifM09} scales in practice. There are however further interesting correspondences and divergences between our method and their's. The first condition that they identified was that if a given data word exits a design at any time, then it must have been captured into the design at same or earlier time. This happens to be equivalent to one of our invariants I4 (though our definition is contra-positive). The second condition they identified is not something we exploited at all in our work as we didn't find it necessary for our approach. This condition says that if one has two copies of the design (RTL) driven by the same control logic and input data stream is identical for all time points other than that one point where a watched data word is captured and if the values at input differ then there must be a difference in the output values at the point where input data streams differ. We fail to see how by proving a property which is based on their second condition would be necessary. In our work we establish one-to-one correspondence between all data words that were accepted at the input interface and the ones that leave the design at the output interface by proving the ordering property which uses all of the input data words as watched words for tracking and ensures that whatever was accepted didn't get lost, corrupted, or duplicated in the design and also remained correctly ordered within the design. If the design had a bug in which it can duplicate the data then there will be at least one instance of a value that will get duplicated and will overwrite the watched data. It will affect the associated tracking counter in that when we expect to see the watched data we will instead see one of the duplicated values appear at the output interface. The positional invariant we used to scale our ordering proofs serves as a useful companion property that gets failed in the presence of bugs in the design. We did not notice any corresponding property in the work done by \cite{DBLP:journals/entcs/BenalycherifM09}. We are also not sure how well their PSL based properties shown for FIFO verification scale as there is no data reported and lastly we are also not sure how much of their formalization was reusable for different designs.
\subsection{Invariants and Assume-Guarantee}
We extended our abstraction model with invariants to accelerate the proof times of the ordering property in many cases making the proof tractable in the first place. For example with the use of positional invariant in case of FIFO verification, we were able to prove  in less than a minute of wall-clock time,  the key ordering and liveness properties even with exponentially increasing depths of the FIFO. The proof of the positional invariant itself scales {\it linearly} as depths of the FIFO increases exponentially. We are not aware of any result from any other work where verification effort and time is scaling linearly (or remains constant) as hardware configuration scales exponentially. 

The performance boost we obtain in our methodology is not only due to using invariants or abstraction but also leveraging the proofs of intermediate invariants using an assume-guarantee flow which we scripted in-house. This is unique to our approach and we are not aware if \cite{DBLP:conf/fmcad/AggarwalCKS11} or \cite{DBLP:journals/entcs/BenalycherifM09} considered this. The striking feature of using invariants in this way is that the process of finding these encourages verification and design engineers alike to consider carefully aspects of micro-architectural specifications which helps build the designs correct in the first place avoiding bugs, or finding more bugs through the use of invariants as explicit checks. In our case one of the first design components that got fixed was a low-power FIFO design where that bug was uncovered through an invariant. The use of assume-guarantee the way we have used it is different from how Ken McMillan~\cite{mcmillan00methodology} used it to cope with circularity and loops in designs. 
\subsection{Scenario splitting abstraction}
The use of scenario splitting is another form of abstraction in that it allows one to focus the verification to be on one scenario at a time and allows us to abstract away the non-interesting scenarios. This is the key to our approach in the verification of memory subsystem arbiter where a combined tracking and observation logic encapsulated in a single model (used in the verification of simple arbiters) is limiting for performance and tractability of the formal proofs. The use of scenario splitting offers a natural abstraction for designs such as these on top of smart-tracker and two-transaction abstraction. We are not aware if anyone else considered this as part of a wider problem reduction strategy. 
\section{Conclusions}\label{concl}
In this paper we presented three key abstractions to address the verification of a range of different designs ranging from simple library components such as FIFOs to simple and complex arbiters, multi-clocked synchronizers and a packet based design used in networking. The use of abstractions together with invariants and assume-guarantee flow provides a powerful methodology to scale and deploy formal hardware verification on industrial strength designs in an automatic manner. We showed how one testbench for FIFO verification using a {\it single ordering assertion check} can verify 100 different types of FIFOs for ordering correctness. Our solution for FIFO verification scales linearly in run time (for positional invariant) or remains constant (for ordering assertion) as design configuration scales exponentially. For arbiter synchronizer, and the packet based design the run time is of the order of seconds. Even for a complex memory subsystem arbiter and a complex packet based design from networking the run time ranges from minutes to an hour. It is not possible to obtain both exhaustive results and find corner case bugs (such as redundant power consumption) with these run times using simulation or emulation. Using a small set of abstractions to verify exhaustively a range of different hardware designs is a completely new way of addressing the challenge of delivering high-assurance products in the market with shrinking time schedules. In the future, we are planning on providing more automation for deducing invariants and to establish that our abstraction is a Galois connection.
\section{Acknowledgment}
We would like to thank John Colley and Neil Dunlop for their time and valuable feedback.  
\bibliographystyle{IEEEtran}
\bibliography{main}
\end{document}